\documentclass[twocolumn,preprintnumbers,superscriptaddress]{revtex4}
\usepackage{epsf}
\usepackage{graphicx}
\usepackage{dcolumn}
\usepackage{bm}
\def\prb{Phys. Rev. B}

\def\be{\begin{equation}}
\def\ee{\end{equation}}
\def\ba{\begin{eqnarray}}
\def\ea{\end{eqnarray}}

\begin{document}

\draft

\title{Universal Aspects of Coulomb Frustrated Phase Separation}
\author{Reza Jamei }
\affiliation{Department of Physics, University of California, Los
  Angeles, California  90095}
  \author{Steven Kivelson }
\affiliation{Department of Physics, University of California, Los
  Angeles, California  90095}

\author{Boris Spivak}
\affiliation{Department of Physics, University of Washington, Seattle, WA 98195}
\

\begin{abstract}
We study the consequences of Coulomb interactions on a system undergoing
 a putative first order phase transition.  In two dimensions (2D),
 near the critical density, the system is universally unstable to the
  formation of new intermediate phases, which
we call ``electronic microemulsion phases,'' which  consist of an intermediate
scale mixture of regions of the two competing phases.  A correlary is that there
can be no direct transition as a function of density from a 2D Wigner crystal to a
uniform electron liquid.  In 3D,
if the strength of the Coulomb
interactions exceeds a critical value, no phase separation occurs, while for
weaker Coulomb strength,  electronic microemulsions are inevitable.
This tendency is considerably more pronounced in anisotropic (quasi 2D or quasi 1D)
systems, where a devil's staircase of transitions is possible.
\end{abstract}


\maketitle

We consider the effect of long range Coulomb interactions on a
system undergoing a first order phase transition between two
compressible states. In the absence of Coulomb interactions, a
first order transition implies an interval of mean density in
which the equilibrium state is macroscopically phase separated
into regions of higher and lower than average density.  A Coulomb
interaction precludes macroscopic phase separation; consequently,
the system can either  undergo a direct, first order phase
transition between the two competing uniform states at the
critical density, or can  form intermediate phases, which we refer
to as ``electronic microemusions,'' which can be thought of as
a mesoscale mixture of the two competing phases. It
has been suggested that  such states resulting from Coulomb
frustrated electronic phase separation are relevant to the physics
of various highly correlated materials, including the cuprate high
temperature superconductors and the collosal magneto-resistance
manganates\cite{physicaC,tokura}, as well as to\cite{spivak,spivakandme} the pure
two dimensional  electron liquid at large $r_s$ (low density).

In the present paper, we consider several universal ({\it i.e.}
independent of microscopic details) aspects of Coulomb frustrated
phase separation.  In two dimensions (2D), we show that even in
the absence of any quenched disorder, a direct first order phase
transition as a function of density between two distinct thermodynamic
phases is forbidden. Rather, in the neighborhood of the putative
critical density, there is an instability of the uniform
state to some form of
mesoscale phase-coexistence, leading to new intermediate phases.\cite{footnote}
 In particular, this theorem applies to the  2D Wigner crystal
to electron fluid transition.\cite{Ciperley}

In 3D, a direct first order transition (with no region of phase
coexistence) occurs if the strength of the Coulomb interaction,
$Q$, exceeds a critical\cite{DeCastro} strength, $Q > Q_c$, while if $Q <
Q_c$, one or more microemulsion phases occur.   Under many
circumstances, at least at mean-field level, the transition from
homogeneous to microemulsion phases as a function of decreasing
$Q$ is a Lifshitz transition, in which the characteristic period
of the mixture diverges as $Q\to Q_c$ from below, and there can be a
Devil's staircase consisting of a complex mixture of commensurate
and incommensurate density waves.  For $Q \ll Q_c$, the
characteristic size of the regions of each phase is small, and
more detailed, microscopic considerations become important.  In
anisotropic (quasi 2D or quasi 1D) systems, $Q_c$ becomes
increasingly large with increasing anisotropy, making some form of
Coulomb frustrated mesoscopic phase separation nearly inevitable.

{\it An explicit model:}  The analysis described in the present
paper is quite general, but it is nevertheless more convenient,
pedagogically, to present it in the context of a simple model
described by the following classical Hamiltonian density \ba {\cal
H} =\sum_{\alpha=1}^D && \frac {\kappa_\alpha} 2 [\partial_a
\phi]^2 +
U(\phi) - \mu(\rho-\rho_c)\phi + \ldots \nonumber \\
&& +\frac 1 2 [\rho -\bar\rho] V [\rho -\bar\rho], \label{H} \ea
where $U$ is a potential, which to be concrete we take to be
$U=\lambda[\phi^2- 1]^2/4$,  $\rho$ is the charge density, $\bar
\rho$ is the (non-dynamical) uniform background charge density,
and we have expressed the Coulomb interaction, in an operator
form, such that $Vf \equiv \int d^Dr^{\prime} V(\vec r -\vec
r^{\prime})f(\vec r^{\prime})$ with $V(r)=Q/r$.  In this model,
the two uniform phases have $\phi= \pm 1+\mu (\bar
\rho-\rho_c)/2\lambda + \ldots$  which we label as $\phi\approx
\pm 1$, assuming $\lambda$ to be large. The  term in ${\cal H}$
proportional to $\mu$ expresses the fact that the $\phi\approx 1$
phase is favored  at higher densities, and the $\phi\approx -1$
phase is favored at low.  For an isotropic  3D system,
$\kappa_\alpha=\kappa$, while we will call a system quasi-2D if
there  exists one direction in which $\kappa_\alpha$ is very
small, and quasi 1D if there  are two such directions.  (We will
label spatial directions in such a way that
$\kappa_1\ge\kappa_2\ge\kappa_3$.)  We use a uniform continuum
notation, but where discreteness is important (such as in a
layered system, in which position in the perpendicular direction
is labeled by a  layer index), it is to be understood
that $\partial_\alpha\phi \equiv [\phi(n+1) - \phi(n)]/a_\alpha$
where $a_\alpha$ is the lattice constant.   The terms $\ldots$
represent higher derivative terms, and terms proportional to
higher powers of $\rho-\bar\rho$.

The significance of $\kappa_\alpha$ and $\lambda$ is that they determine the local
({\it i.e.} the $Q\to 0$ limit) structure of an interface between the two phases
running perpendicular to the $\alpha$ direction:  Specifically,  the surface
tension is $\sigma_{\alpha}=(\pi/2)\sqrt{\lambda\kappa_\alpha/2}$ and the width of
the
interface is $a_{\alpha}=\sqrt{\kappa_{\alpha}/2\lambda}$.

Because the Hamiltonian density is a quadratic function of $\rho$, density
fluctuations can be 
integrated  out to yield an effective Hamitlonian for $\phi$,
alone.  The result can be expressed formally as
\ba
{\cal H} ^{eff}=\sum_{\alpha=1}^D && \frac {\kappa_\alpha} 2 [\partial_a \phi]^2 +
U(\phi) - \mu(\bar\rho-\rho_c)\phi + \ldots \nonumber \\
&&- \frac 1 2 \sum_{\alpha=1}^D \partial_\alpha \phi \tilde V \partial_\alpha \phi
,
\label{Heff}
\ea
where the renormalized interaction, $\tilde V$, is defined in terms of the  inverse of $V$ according
to
$-\nabla^2\tilde V = \mu^2V^{-1}$, or taking the Fourier transform:
$\tilde V_{\vec k}=\mu^2[k^2V_{\vec k}]^{-1}$.

{\it  The 2D case:}  In 2D, the Fourier transform of $\tilde V$ is $\tilde V_{\vec 
k}=
\mu^2 [\pi Q |k|]^{-1}$. 
Transforming back to real space, the
final term in Eq. \ref{Heff} is
\ba
E_{Coulomb} = -\frac {\mu^2} {4\pi^2Q} \int d^2rd^2r^{\prime} \frac {\vec\nabla
\phi(\vec r)\cdot\vec \nabla\phi(\vec r^{\prime})}{|\vec r - \vec r^{\prime}|}
\label{EC}
\ea

In the limit in which the interfaces between phases are narrow
compared to the size of domains, one can neglect the gradient
of $\phi(r)$ in the bulk of the phases, so the expression for
the total energy can be simplified to read 
\ba 
E = \mu[\bar \rho -
\rho_c][A_+ - A_-] + \int d\vec l \sigma(\theta) - \frac {\mu^2}
{\pi^2Q} \int \frac {d\vec l \cdot d\vec l^{\prime} } {|\vec l
-\vec l^{\prime}|}, 
\label{strings} 
\ea where the integrals run
along the interface in the direction such that the $\phi\approx
+1$ phase is to the left, $A_{\pm}$ are the areas of the two
phases, $\sigma(\theta)$ is the microscopic contribution to the
surface tension, which can depend on the direction of the
interface, and any short-distance singularities in the final
integral are cutoff at the scale of the interface width, $a$.

Eqs. \ref{EC} and \ref{strings} are the principle results of our paper;  the classical low 
temperature thermodynamics is obtained by integrating over all domain patterns with Boltzman weight
determined by this effective Hamiltonian.  The key feature of this sum is that the third term in Eq.
\ref{strings} is negative and logarithmically divergent whenever the a domain is large. 
Thus, close enough to the point of the phase transition,
micro-emulsion phases which are phase separated on a mesoscopic
scale always have lower energy than the uniform phases.\cite{fisher}
It has previously been shown\cite{brazovski} 
 on quite general
grounds that if there is a direct
transition from a crystal to a uniform fluid, it must be first order. Combined with the present
result, this  implies that there can be no direct transition.  Rather, there must exist
one or more intermediate phase, and a sequence two or more
continuous transitions\cite{KT}. Moreover, in the case of the the 2DEG, 
the first of these transitions
must occur at strictly higher density (smaller $r_s$) than the
putative  Wigner  crystal to uniform liquid transition.\cite{CiperleySim}

To find the mean-field phase diagram one should minimize Eq. \ref{strings} with
respect to the geometry of the interfaces.
 Remarkably,  the same expression for the
interfacial energies (the third term in Eq. \ref{strings}) arises
\cite{Marchenko,vanderbilt,spivak,spivakandme} in a 2D problem
where phase separation is frustrated by long-range dipolar
interactions (although in this case the first,  bulk contribution
to the energy is different). Thus what is known about
that problem can be easily applied to the present case.

On the mean field level the shape of the micro-emulsion depends of
a degree of anysotropy of the surface tension. The simplest
situation arises if $\sigma$ is highly anisotropic - in this case,
the domain walls always lie preferentially along the easy
direction and thus, at least at mean-field level, the lowest
energy microemulsion phases are striped \cite{spivakandme}.

The energy per unit area, relative to the uniform phase, of an alternating
array of
stripes of $\phi\approx +1$ of width $W$ and $\phi\approx -1$ of
width $L-W$ is computable from Eq. \ref{strings}: 
\ba
\varepsilon  =2L^{-1}\left\{\mu\ \delta\bar\rho\  W + \sigma \left[1
-\gamma\log(W/a) -\gamma f(W/L)\right ] \right\}
\label{stripes} 
\ea
where $\delta\bar\rho = \bar\rho-\rho_c$, $\gamma \equiv \mu^2/(\pi^2\sigma Q)$, and $f(x)=
\log[\sin(\pi x)/\pi x]$.
For $\bar \rho=\rho_c$, this
energy is minimized by alternating stripes of equal width, $L=2W$,
with $W=(2/\pi)W_0$ and \ba W_0=a\exp[1+1/\gamma] \ea
 is a characteristic emergent length scale.  For
$\bar \rho > \rho_c$, the ratio of $W/L$ decreases monotonically,
until as 
$\delta\rho \to \Delta \rho
\equiv (\sigma\gamma/\mu)W_0^{-1}$, the period of the
stripes diverges as 
\be 
L \sim
W_0\sqrt{\frac{3\zeta(2)\Delta\rho}{\Delta\rho-\delta\bar\rho}}
\ee and the width of the stripes, $W \to W_0$, where
$\zeta(2)=\pi^2/6$ is the zeta function.

Thus, as a function of decreasing $\bar\rho$, there is a
``Lifshitz'' transition \cite{LifshitzComment} from the uniform
phase to the stripe phase at $\bar \rho =\rho_c+\Delta\rho$, then
a continuous evolution of the stripe phase with the identity of
the minority and majority phases interchanging at
$\bar\rho=\rho_c$, and finally a second Lifshitz transition at
$\bar\rho=\rho_c -\Delta\rho$.  The resulting mean feild phase
diagram   is qualitatively identical to the phase diagram in the
case with dipolar interactions -- see Fig. 2c of Ref.
\cite{spivakandme}. 
However, in the dipole case, the density differences between the 
two phases is roughly determined by a Maxwell construction, and the characteristic stripe width is
determined by the  size of the dipole, $d$,
 which is an independent physical parameter, which in some physical realizations can be large
compared to the spacing between particles.  In the present case, the
density contrast and the characteristic width of the stripes, $W_0$, is
determined by microscopic physics. 

When the surface tension is more nearly isotropic, the mean-field
phase diagram is more complicated.
Although for $\bar\rho$ near $\rho_c$, stripe
phases continue to have the lowest energy, near the transition to
the uniform phase, bubble phases, consisting of isolated bubbles
of the minority
phase, have lower energy than
the stripe phase.  The Lifshitz points at $\bar \rho=\rho_c\pm
\Delta\rho$ now mark the transition between a uniform and a
long-period bubble phase \cite{spivak,spivakandme}. In the case of
the triangular Wigner crystal-Fermi liquid transition the anisotropy of the surface
tension is such as to produce stable  bubble phases.  
Although the optimal bubble is facetted,
one can establish the stability of the bubble phase by using as a variational
ansatz hexaganol bubbles.
On approach to the Lifshitz point, the size of the bubbles is of order $W_{0}$ while the distance
between them diverges as
\begin{equation} L_{B} \sim
W_{0}[\Delta \rho/(\Delta
\rho-\delta\bar{\rho})]^{-\frac{1}{3}}.
\end{equation}
This is not quite the end of the story; since the transition
between the bubble and stripe phases is first order, a second
level of frustrated phase separation occurs producing a new set of
intermediate phases. These states consist of alternating stripes
of  stripes and stripes of bubbles. The resulting phase diagram is
qualitative identical to that shown in Fig. 2d of Ref.
\cite{spivakandme} for the 2D dipolar case. Again, the difference is
that in the dipolar case the size of the bubbles is proportional
to the size of the dipole, and can be parametrically large.

Because of the importance of the 2D case, it is worth deriving Eq.
\ref{strings} in a second way. For simplicity, consider the
situation in which the background charge density is tuned to the
critical value at which the putative first order transition
occurs, $\bar\rho =\rho_c$.  We compute the difference in energy
between the uniform density state
 and that in which there is an interface such that (other than in a narrow
interface region of width $2a$), there is phase $\phi\approx-1$ for $x <
-a$ and phase $\phi\approx +1$ for $x  > a$.  So as to minimize the
Coulomb cost of this interface, while at the same time gaining
maximal energy from phase separation, we allow the charge density
to assume the profile $\rho(\vec r) = \bar \rho + \phi(x)
\Delta(a|x|)^{-1}$. (This expression is identical to that arising
in the problem of  a contact potential of two metals with
different work functions \cite{LandauLifshitz}.) With this profile,
we easily see that the Coulomb cost of the interface is $E_c =
\Delta^2 Q L\log(W/a)$, where $L$ is the length of interface and
$W$ is the transverse width of the two regions separated by it.
Similarly, the energy of phase separation gained within each
region by the density redistribution is $E_{PS}= - 2\mu \Delta
(L/a) \log(W/a)$. Because the major contributions to these
logarithms comes from regions far from the interface, where the
deviations from uniform density are small and slowly varying, the
result is insensitive to the inclussion of any further terms in
the energy functional. The energy of this interface is minimized
by $\Delta = \mu/aQ$, at which point
 \be
 E_{interface} = [\sigma_0 - (\mu^2/aQ) \log(W/a)] L
 \label{interface}
 \ee
 which is manifestly negative for large enough $W$.

 Taking, as an example, a striped state, where $W$ is the width of the stripes,
the logarithm in Eq. \ref{interface} can be seen to be precisely the same one that
comes from integrating the expression in Eq. \ref{strings} over the interface, Eq.
\ref{stripes}.
Eq. \ref{interface}, by itself, constitutes a proof that a first
order phase transition is forbidden - it implies an absolute
instability of any uniform (or macroscopically phase separated)
state at the point of the putative transition.  Although the analysis that leads to this
conclussion is classical,
because it is a long-wave-length instability, it is unaffected by quantum fluctuations,
and so applies at zero temperature, $T=0$, as well as non-zero $T$. 

The characteristic size of bubbles and stripes depends
exponentially on the ratio of microscopic energies, $\gamma$:  If
$\gamma$ is roughly 1 or more, then the size of the domains
 is microscopic.
 Moreover, in this case, there is no reason to expect quantum or
thermal  fluctuations to be small.  Consequently, although the
instability of the uniform phases is a robust long-distance phenomenon,
the mean-field  phases we have found may
or may not survive these fluctuations.  However, because of the
exponential dependence, it can happen that $\gamma$ is small
compared to 1, in which case $W_0$ can be very large compared to
$a$.  In this case, the mean- field analysis presented here should
be reliable, and $W_0$ is a non-trivial emergent length. The
significance of the classical and quantum fluctuations has been
discussed in \cite{spivak,spivakandme}.

{\it The 3D case:}
In 3D, it follows simply from the fact that the Fourier transform of the Coulomb
interaction is $V(\vec k) = 4\pi Q/k^2$, ${\cal H}^{eff}$ is
local
\be
{\cal H}^{eff} =\sum_{\alpha=1}^3 \frac {[\kappa_\alpha - \mu^2/4\pi Q]} 2 
[\partial_a
\phi]^2 +
U(\phi) - \mu\bar\rho\phi  + \ldots
\label{3DH}
\ee
Thus, if the Coulomb interactions are sufficiently strong that $ Q > A_c \equiv 
\mu^2/
4\pi\kappa_3$,
then as a function of $\bar\rho$, the system undergoes a direct first order
transition from a uniform $\phi\approx 1$ phase for $\bar\rho > \rho_c$, to a 
uniform
$\phi\approx -1$ phase for
$\bar\rho < \rho_c$.  In contrast to 2D, there is no absolute proscription against
first order transitions in 3D.

However, for weaker Coulomb interactions, for $\bar \rho$ in the
neighborhood of $\rho_c$, there is an intermediate modulated phase
whose precise structure is determined by the higher derivative
terms which are not explicitly exhibited. (In part, this case has been
previously considered in \cite{DeCastro}.) For illustration, consider the phase
diagram at the critical density, $\bar \rho=\rho_c$.  At the
critical value, $Q=Q_c$, the coefficient of the leading stiffness
term changes sign, the uniform state must be unstable, and higher
order elastic constants (more derivatives) must be included in the
effective Hamiltonian. Most simply, we can include the next order
terms in Eq. \ref{3DH}, $\ldots = \sum_\alpha\kappa_\alpha^\prime
(\partial^2_\alpha \phi)^2 +\ldots.$  So long as these higher
order elastic constants remain positive, the transition to the
modulated phase as a function of $Q$ is the classic Lifshitz
transition.  Specifically, as $Q$ approaches the critical value
$Q_c$ from below, the period of the modulated phase diverges as $L
\sim a\sqrt{Q_c/(Q_c-Q)}$. Alternatively, if 
$\kappa_3^{\prime} <0$, this Lifshitz transition is preempted by a
first order transition (at a larger critical $Q$) to a modulated
phase with a short period.

{\it Layered systems:}  A natural realization of a quasi 2D system is in a layered
material.  This is still a 3D system, and so subject to the analysis of
the previous paragraphs, but now there can be significant effects of the discrete,
lattice structure.  In particular, so long as the spacing between layers, $a_3$,
is not too large, this discreteness introduces the only important modification of
the above discussion.  Where the continuum theory predicts a discontinuous 
transition
to the modulated phases, lattice discreteness generally locks the period of
the resulting modulation to a specific, low-order commensurate structure.
Alternatively, near the Lifshitz transition, the competition between the period
favored by the elastic constants and the underlying lattice typically results in a
Devil's staircase of modulated phases, and moreover can lead to intrinsically
glassy behavior associated with the pinning of domain walls.

The easiest way to get a flavor of the resulting possibilities is by simply
discretizing the spatial derivatives in Eq. \ref{3DH}:  in the case in which
$\kappa_1=\kappa_2  > \kappa_3$, and $\kappa^{\prime}_\alpha >0$, the resulting
model is the soft-spin version of the much studied\cite{ANNNI} antiferromagnetic
next-nearest-neighbor Ising (ANNNI) model, in which the antiferromagnetic coupling
between second neighbor planes is proportional to $\kappa_3^\prime$, the couplings
between nearest-neighbor planes has a strength (and sign) which varies as a
function of $\kappa_3 -\mu^2/4\pi Q$, and  the in plane nearest-neighbor
ferromagnetic couplings are proportional to $\kappa_1$.  This model has a
remarkably subtle and beautiful phase diagram consisting of uniform
(ferromagnetic) phases, short period commensurate phases, of which the most
prominent are the period 2 (alternating up and down planes) and period 4 (2 up 2
down) phases, and then a ``Devil's flower'' consisting of high order commensurate,
and incommensurate phases.  Presumably, the more subtle aspects of this
phase diagram are lost when thermal or quantum fluctuations are included, but the
tendency for Coulomb frustrated phase separation to produce  patterns of
alternating high and low density layers is  robust.

In the limit of larger $a_3$, the phase diagram is still more complex.  So long as
$a_3$ is less than the characteristic size, $W_0$, which characterizes the domain
size in an isolated layer, interlayer phase separation is  energetically
preferred over intra-layer phase separation.  However, for $a_3 \gg W_0$, 2D
patterns of phase separation dominate the physics at shorter length scales, while
intralayer considerations become important only at much longer distances.

Quenched disorder produces its own form of locally pinned mixtures
of the two competing phases.  Distinguishing the two effects in experiment
requires studying the properties of progressively cleaner systems.
With this
caveat, we believe the present results are significant for a host of 
phenomena in layered and quasi-1D crystalline materials\cite{2DEGexp}, as well as
for\cite{spivak,spivakandme} the 2DEG in high mobility
semiconductor devices at large $r_s$.


We acknowledge useful discussions with S.Chakravarty and D.Fisher.
This work was supported in part by the National Science Foundation under Contracts 
No.
DMR-01-10329 (RJ and SAK) and DMR-0228014 (BS).


\begin{thebibliography}{99}
\bibitem{footnote} This theorem is similar to a result, obtained in Refs.
\cite{spivak,spivakandme,Marchenko} concerning the non-existence of first order transitions in 2D 
systems with repulsive, dipolar interactions.  More generally, the present theorem can be
straightforwardly generalized to $D$ dimensional systems with
repulsive interactions which decay as $V(r) \sim 1/r^{\alpha}$
with $D-1 \le \alpha \le D$, while the proof for the dipolar case
applies so long as $D < \alpha \leq D+1$.
\bibitem{Ciperley} We note that variational Monte-Carlo calculations
which have previously
 determined\cite{CiperleySim} a critical value of $r_s=38$ for Wigner crystallization
in 2D are based on the now falsified assumption that there is a
direct Wigner crystal-Fermi liquid transition. We also note numerical simulations in Ref.
\cite{Pichard} were interpretted as evidence of the existence of intermediate
phases.
\bibitem{fisher}  Note:  there could exist special circumstances in which $\mu=0$, where higher  powers must be taken into account in Eq. (1).  In this case, our proof is not applicable.
\bibitem{KT}  The Brazovski argument\cite{brazovski} applies directly at $T=0$, where the Wigner crystal
is long-range ordered, and the Berezinski-Kosterlitz-Thouless-Halperin-Nelson (BKTHN) transition is impossible.  At $T>0$, where the Wigner crystal has only quasi-long-range order,
the BKTHN scenario of a sequence of two, continuous transtions with an intermediate hexatic
phase competes with a first order direct transition.  The present analysis yeilds the stronger conclussion that in the Coulomb case, a first
order transition is forbidden, and at low enough $T$, the implied intermediate phases 
must be the finite $T$ decendents of the $T=0$ microemulsion phases.
\bibitem{LifshitzComment} A  Lifshitz point
is conventionally characterized by vanishing amplitude of the order parameter and
diverging period of the superstructure. We use the same term, here,
for a related, but somewhat different type of transition in which
the distance between stripes (or bubbles) diverges
at the critical point, while the size of the bubbles and the
density profile inside them remains constant.
\end{thebibliography}
\end{document}